\documentclass[aps,showpacs]{revtex4}
\usepackage{epsfig}
\begin{document}

\title{Best linear forecast of volatility in financial time series}
\author{M. I. Krivoruchenko}
\affiliation{Institute for Theoretical and Experimental Physics$\mathrm{,}$ B. Cheremushkinskaya 25\\ 117259 Moscow, Russia}
\affiliation{Institut f\"{u}r Theoretische Physik$\mathrm{,}$ Universit\"{a}t T\"{u}bingen$\mathrm{,}$ Auf der Morgenstelle 14\\ D-72076 T\"{u}bingen, Germany}
\affiliation{Metronome-Ricerca sui Mercati Finanziari$\mathrm{,}$ C. so Vittorio Emanuele 84\\ 10121 Torino, Italy}

\begin{abstract}
The autocorrelation function of volatility in financial time series is fitted well by 
a superposition of several exponents. Such a case admits an explicit analytical 
solution of the problem of constructing the best linear forecast 
of a stationary stochastic process. We describe and apply the proposed analytical 
method for forecasting volatility. The leverage effect and volatility clustering are 
taken into account. Parameters of the predictor function are determined numerically for 
the Dow Jones 30 Industrial Average. Connection of the proposed method to the 
popular ARCH models is discussed.
\end{abstract}
\pacs{89.65.Gh, 89.75.Da, 02.50.Ng, 02.50.Sk}

\maketitle

\section{Introduction}

The methods developed in studying complex physical systems have been
successfully applied throughout decades to analyze financial data \cite
{bachelier,levy,mandelbrot}. The quantitative study of financial data
continue to attract the growing interest motivated by the existence of
universal features in the dynamics of different markets, such as power-law
tails of the return distributions \cite
{clark,mantegna,geman,para,manst,and,ple,derman}, scaling as a first
approximation \cite{mandelbrot} and deviations from scaling of the empirical
return distributions \cite{cont,mantegna,manst}, volatility clustering \cite{ding,cont}%
, and leverage effect \cite{black,cox,christie}. The phenomenological and
microscopic models \cite{manst,and,ple,derman,mik,mikw,jpb,holyst} have been proposed to
explain the established stylized facts. The field of research connected to modeling
financial markets has been named Econophysics.

A stock's volatility represents the simplest measure of its riskiness or
uncertainty. Formally, the volatility is the annualized standard deviation
of the stock's returns during the period of interest. The random walk model
proposed by Bachelier in 1900 year \cite{bachelier} presupposes a constant
volatility. There is an ample empirical evidence, however, that
the volatility is not a constant, but represents a random variable. Two well
established stylized facts concerning the volatility are long ranged
volatility-volatility correlations that are also known as volatility
clustering \cite{ding} and return-volatility correlations that are also
known as leverage effect \cite{black,cox}.

The volatility is a key variable to control risk measures associated with
the dynamics of prices of financial assets. The implied volatility extracted
from options prices represents a market estimate of future volatility. A
pure exposure to future volatility is provided by the volatility swaps \cite
{swaps,carr}. The volatility enters all options pricing models, so its
knowledge has a great value for estimate of the equilibrium options
state-price distributions.

The volatility clustering manifests itself in the occurrence of large
changes of the index at neighboring times (observed localized outbursts).
The leverage effect has its origin in the observed negative correlation
between the past returns and future volatility. The possible explanation to
this effect \cite{black,cox,christie} is due to the fact that negative
returns increase financial leverage and extend the risk for investors and
thereby a stock's volatility. A statistical study \cite{bou} demonstrates
clearly that the leverage effect is one-directional: past returns correlate
with future volatility only.

In this paper, we propose an analytical method to evaluate future volatility
as a linear function of the lagged volatility and lagged returns. The method
takes the volatility clustering and leverage effect into account and
provides for stationary stochastic processes the smallest forecasting error
in the class of all linear functions. In this precise sense, we talk on the
best linear forecast (BLF) of the volatility.

The BLF problem for a stationary stochastic process was formulated
and solved by Kolmogorov \cite{kolm} in 1941 year and Wiener \cite{wien}
in 1949 year. A modern review of the BLF methods can be found in Ref. \cite
{portenko}. We apply these methods to construct the BLF volatility function
for the Dow Jones 30 Industrial Average (DJIA).

The outline of the paper is as follows: In the next Sect., we remove the
leverage effect from the original time series to work with a reduced
volatility $\chi (t)$ that has by definition a vanishing covariance with the
past returns. The spectral density of a stochastic process can be
factorized, $f(\omega )=|\varphi (\omega )|^{2},$ if its correlation
function represents a superposition of the exponential functions. An
explicit expression is derived for the amplitude $\varphi (\omega )$. The
analytical properties of the amplitude $\varphi (\omega )$ in the complex 
$\omega $-plane are important to provide an explicit representation of the predictor
function. In Sect. 3, the BLF problem is analyzed further to account for the
reduced volatility clustering and to construct the BLF function. In Sect.
4, we fit 100+ years of data of the daily historical volatility of the DJIA
in order to determine parameters of the BLF function. Numerical estimates
are given to illustrate the developed method. The minimization of the
forecasting error for the reduced volatility predictor function 
is shown to be
equivalent to the minimization of the forecasting error of the original
volatility time series. An explicit expression for the forecasting error is
given. In Conclusion, a connection of the BLF method with the ARCH models 
\cite{engle,arch1,arch2,arch3}, in which future variance is also represented as a
linear combination of the past observables, is discussed.

\section{Factorization of spectral density}

The evolution of a market index value
or a stock price $S(t)$ is described by equation (see e.g. \cite{hull}): 
\begin{equation}
\frac{dS(t)}{S(t)}=\mu dt+d\psi (t).  \label{dS/S}
\end{equation}
The value $d\psi (t)$ is a noise added to the path followed by $S(t),$ with
the expectation value $\mathrm{E}[d\psi (t)]=0$ and the variance of $\mathrm{%
Var}[d\psi (t)]=\sigma (t)^{2}dt.$ The volatility $\sigma (t)$ represents a
generic measure of the magnitude of market fluctuations. We consider a
discrete version of the random walk problem by setting $dt=1$, $d\psi
(t)=\xi (t),$ and $dS(t)=S(t)-S(t-1).$ The sampling intervals are enumerated
by integer time parameter $t$.

The volatility $\sigma (t)$ is a hidden variable and its extraction
form the market observables is a separate difficult task. The possible
estimator $\eta (t)=|\xi (t)|$ of the volatility is defined in terms of
returns 
\begin{equation}
\xi (t)=(S(t)-S(t-1))S(t)^{-1}-\mu .  \label{defreturn}
\end{equation}
In what follows, the term ''volatility'' refers to the estimator $\eta
(t)=|\xi (t)|$, the annualizing factor will not apply. A use of the variance 
estimator $v(t)=|\xi (t)|^{2}$ would complexify the problem due to divergences connected
to the existence of power-law tails (''variance of variance'' is infinite, 
\textrm{Var}$[\xi ^{2}]=\infty ,$ since $dF(\xi )\sim d\xi /\xi ^{4}$ at $%
\xi \gg 1,$ see e.g. \cite{para}). At large time scales, different
estimators are expected to be close to volatility $\sigma (t)$ and to each
other. The problem of efficiency of volatility estimators is postponed for
other studies.

It is usually assumed that financial time series constitute stationary
stochastic processes the autocorrelation functions of which depend on the
relative time only. The stock evolution problem is assumed therefore to be
invariant with respect to time translations.

First, we remove from the time series $\eta (t)$ the leverage effect using
the variable $\chi (t):$%
\begin{equation}
\chi (t)=\eta (t)-\sum_{s}\mathrm{Cov}[\eta (0),\xi (-s)]\mathrm{Var}%
^{-1}[\xi ]\xi (t-s).  \label{mod}
\end{equation}
The decomposition (\ref{mod}) has a predictive power, since $\mathrm{Cov}%
[\eta (t),\xi (t-s)]\sim \theta (s)$, so $\chi (t)$ depends on the lagged
price increments only. Note that $\mathrm{E}[\eta ]=\mathrm{E}[\chi ],$
since $\mathrm{E}[\xi ]=0$. Due to the definition (\ref{mod}) and in virtue
of equation 
\begin{equation}
\mathrm{Cov}[\xi (t),\xi (s)]=\delta _{ts}\mathrm{Var}[\xi ],  \label{corr1}
\end{equation}
that holds true for sampling intervals greater than 20 Min$.$ \cite{cont,para},
we have 
\begin{equation}
\mathrm{Cov}[\chi (t),\xi (s)]=0.  \label{corr2}
\end{equation}
The reduced volatility $\chi (t)$
does not experience the leverage effect. So, its predictor depends on the past 
$\chi (t)$ only. It is possible therefore to focus
on the volatility clustering only, while the leverage effect is taken into
account explicitly through Eq.(\ref{mod}). 

The autocorrelation function $B(t)=\mathrm{Cov}[\chi (t),\chi (0)]/\mathrm{E}%
[\chi ^{2}]$ of a stationary stochastic process $\chi (t)$ can
be fitted in many cases by a superposition of exponents 
\begin{equation}
B(t)=\sum_{i=1}^{n}d_{i}e^{-\alpha _{i}|t|}.  \label{corr}
\end{equation}
The best linear forecast of the observable $\chi (t)$ in such a case
simplifies substantially. 
The case $n=1$ is discussed in Ref. \cite{portenko}. We provide
a solution of the BLF problem for arbitrary values of $n$.

The spectral density $f(\omega )$ of the stochastic process $\chi (t)$ is
given by the Fourier transform of the correlations coefficient (\ref{corr}): 
\begin{equation}
f(\omega )=\sum_{t=-\infty }^{+\infty }e^{-i\omega t}B(t)=\sum_{i=1}^{n}d_{i}%
\frac{1-e^{-2\alpha _{i}}}{(1-e^{-\alpha _{i}}u)(1-e^{-\alpha _{i}}\frac{1}{u%
})}  \label{four}
\end{equation}
where $u=\exp (-i\omega ).$ The function $f(\omega )$ can be represented in
the form 
\begin{equation}
f(\omega )=P_{n-1}(\phi )\left( \prod_{i=1}^{n}(1-e^{-\alpha
_{i}}u)(1-e^{-\alpha _{i}}\frac{1}{u})\right) ^{-1}.  \label{four1}
\end{equation}
where $\phi =\frac{1}{2}(u+\frac{1}{u})$ and 
\begin{eqnarray}
P_{n-1}(\phi ) &=&2^{n}\exp (-\sum_{i=1}^{n}\alpha
_{i})\sum_{i=1}^{n}d_{i}\sinh (\alpha _{i})\prod_{k\neq i}^{n}(\cosh (\alpha
_{i})-\phi )  \nonumber \\
&=&D_{n}2^{n-1}\exp (-\sum_{i=1}^{n-1}\nu _{i})\prod_{i=1}^{n-1}(\cosh (\nu
_{i})-\phi ).  \label{poly}
\end{eqnarray}
The amplitude $\varphi (u)$ such that $f(\omega )=\varphi (u)\varphi (u)^{*}$
can be chosen to be analytical, rational and regular at $|u|<1$: 
\begin{equation}
\varphi (u)=D_{n}^{1/2}\prod_{i=1}^{n-1}(1-e^{-\nu _{i}}u)\left(
\prod_{i=1}^{n}(1-e^{-\alpha _{i}}u)\right) ^{-1}.  \label{phi}
\end{equation}
The additive representation 
\begin{equation}
\varphi (u)=\sum_{i=1}^{n}c_{i}\frac{1}{1-\beta _{i}u}  \label{phi additive}
\end{equation}
is completely equivalent to the multiplicative representation (\ref{phi}).
Here, $\beta _{i}=e^{-\alpha _{i}}$ and 
\begin{eqnarray}
D_{n} &=&2\exp (-\sum_{i=1}^{n}\alpha _{i}+\sum_{i=1}^{n-1}\nu
_{i})\sum_{i=1}^{n}d_{i}\sinh (\alpha _{i}),  \nonumber \\
c_{i} &=&D_{n}^{1/2}\prod_{k=1}^{n-1}(e^{-\alpha _{i}}-e^{-\nu _{k}})\left(
\prod_{k\neq i}^{n}(e^{-\alpha _{i}}-e^{-\alpha _{k}})\right) ^{-1}, 
\nonumber \\
\cosh (\nu _{i}) &=&\phi _{0}^{i}  \label{parameters}
\end{eqnarray}
where $\phi _{0}^{i}$ are $n-1$ roots of equation $P_{n-1}(\phi _{0}^{i})=0.$
For $n=1$, $c_{1}=d_{1}^{1/2}\sqrt{1-\beta _{1}^{2}}.$ The analytical
solutions for $\nu _{i}$ exist up to $n=5$.

The knowledge of the Fourier transform
of the autocorrelation function is not sufficient for a complete
reconstruction of the Fourier transform of the
stochastic process $\chi (t)$. If the autocorrelation function
represents a superposition of the exponents (\ref{corr}), the problem admits
a solution $\varphi (u)$, such that $f(u) = |\varphi (u)|^2$, in the class of
rational functions. If we require further that the function $\varphi (u)$ be
regular at $|u| \leq 1$, an unambiguous solution $\varphi (u)$ can be provided.
This solution coincides with the Fourier transform of the time series $\chi (t)$ up to a phase
factor. It is remarkable, that we need not to know the phase, since all the
relevant information is contained in the spectral density $f(\omega )$. The 
BLF problem simplifies then considerably due
to the special analytical properties of the function $\varphi (u)$.

\section{BLF function}

The correlation function corresponding to the spectral density (\ref{four}) 
can be found from the inverse Fourier transform. In case of (\ref{phi}), 
we consider first $t>0$: 
\begin{equation}
B(t)=\int_{-\pi }^{\pi }e^{i\omega t}\varphi (\omega )\varphi (-\omega )%
\frac{d\omega }{2\pi }=-\int_{\mathcal{C}_{r}}\frac{1}{u^{t+1}}\varphi
(u)\varphi (\frac{1}{u})\frac{du}{2\pi i}
\label{invft}
\end{equation}
where $\mathcal{C}_{r}=\{e^{-i\omega },\omega =-\pi ...\pi \}.$ The poles of 
$\varphi (u)\,$are located at $|u| \geq R = \mathrm{min} \{ e^{\alpha_{i}}\}$, 
while the poles of $\varphi (\frac{1}{%
u})\,$ are located at $|u| \leq 1/R$. We move the contour $\mathcal{C}_{r}$ to
infinity and get
\begin{equation}
B(t)=\sum_{i=1}^{n}\sum_{k=1}^{n}\beta _{i}^{|t|}\frac{c_{i}c_{k}}{1-\beta
_{i}\beta _{k}}.  \label{newcorr}
\end{equation}
Comparison with Eq.(\ref{corr}) gives 
\begin{equation}
d_{i}=\sum_{k=1}^{n}\frac{c_{i}c_{k}}{1-\beta _{i}\beta _{k}}.  \label{di}
\end{equation}
The same result (\ref{newcorr}) comes out at $t<0$. For $n=1,$ $B(t)=\beta
_{1}^{|t|}c_{1}^{2}/(1-\beta _{1}^{2}).$

A stochastic process $\chi (t)$ can be represented as a linear combination
of a normally distributed uncorrelated sequence $\zeta (t)$ $\sim N(0,\sigma _{\chi })$
with $\sigma _{\chi }^{2}=\mathrm{E}[\chi ^{2}],$%
\begin{equation}
\chi (t)=\mathrm{E}[\chi ] + \sum_{s=0}^{+\infty }C(s)\zeta (t-s),  \label{firepr}
\end{equation}
provided that the spectral function admits the factorization and the
amplitude $\varphi (u)$ is regular at $|u| = 1<R$ (see e.g. \cite{portenko}%
). The expansion coefficients equal 
\begin{equation}
C(t)=\sum_{i=1}^{n}c_{i}\beta _{i}^{t}.  \label{ficoeff}
\end{equation}
It is remarkable that only retarded $\zeta (s)$ enter the
summation in Eq.(\ref{firepr}). This is a consequence of the convergence
of the Taylor expansion of the amplitude $\varphi (u)$ at $|u| = 1$, which is
in turn a consequence of the analyticity at $|u|<R$: The convergence radius
of the expansion is associated with the first pole at $|u_{1}|=R$. 
The stationary stochastic process $\chi(t)$ can be interpreted as a result
of filtering the normal sequence $\zeta (t)$.

\begin{figure}[tb]
\vspace{2cm}
\par
\begin{center}
\leavevmode
\epsfxsize = 12cm \epsffile[25 60 573 465]{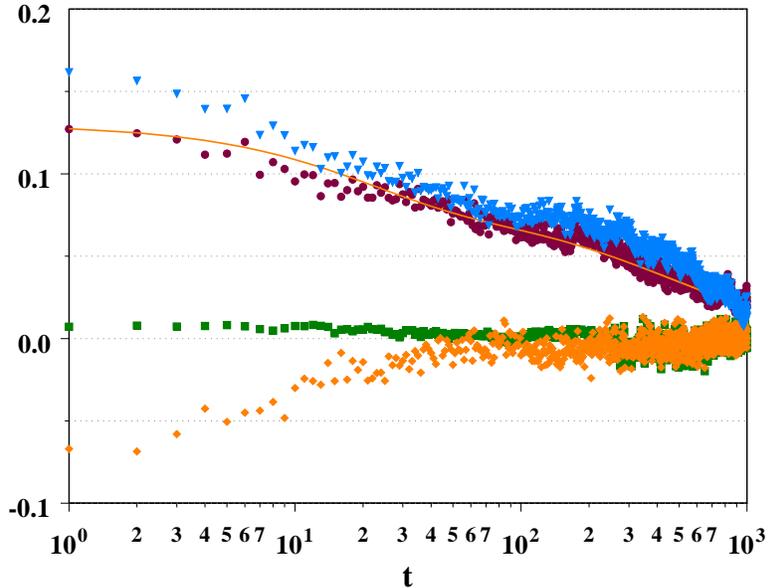}
\end{center}
\caption{ Empirical correlation coefficients \textrm{Corr}$[\eta (t),\eta
(0)],$ \textrm{Corr}$[\eta (t),\xi (0)]$ and \textrm{Corr}$[\chi (t),\chi
(0)],$ \textrm{Corr}$[\chi (t),\xi (0)]$ versus the number of trading days $%
t $ of the Dow Jones 30 Industrial Average$.$ The reduced volatility $%
\chi (t)$ is defined through Eq.(\ref{mod}). The correlation coefficients
are calculated using 100+ years of the daily quotes, starting on the May 26, 1896 
and ending on the December 31, 1999 (i.e. a total of 28507
trading days). The values
of \textrm{Corr}$[\eta (t),\eta (0)]\ $and \textrm{Corr}$[\eta (t),\xi (0)]$
are denoted, respectively, by triangles and diamonds. The values
of \textrm{Corr}$[\chi (t),\chi (0)]\ $and \textrm{Corr}$[\chi (t),\xi (0)]$
are denoted by circles and boxes. The solid curve is the
exponential fit (\ref{corr}) with parameters given in Table 1. It is seen
that \textrm{Corr}$[\chi (t),\xi (0)]$ $\approx 0.$ The leverage effect is
thus removed from $\chi (t).$ }
\label{fig4}
\end{figure}

The BLF function for the time horizon $\tau $ has the form \cite
{portenko}
\begin{equation}
\hat{\chi}_{\tau }(t)=\mathrm{E}[\chi ]+\sum_{s=\tau}^{+ \infty }\frac{\Xi _{\tau
}(0)^{(s)}}{s!}(\chi (t-s)-\mathrm{E}[\chi ]).
\label{xihat}
\end{equation}
The weight coefficients $\Xi _{\tau }(0)^{(s)}$ are derivatives of the
function $\Xi _{\tau }(u)=\varphi _{\tau }(u)/\varphi (u)$ at $u=0$. Here, 
\[
\varphi _{\tau }(u)=\sum_{s=\tau }^{+\infty
}C(t)u^{s}=\sum_{i=1}^{n}c_{i}(\beta _{i}u)^{\tau }\frac{1\ }{1-\beta _{i}u}.
\]
For constructing a linear prognosis function, the overall normalization factor in $B(t)$ 
is not important, since it drops out from the ratio $\varphi _{\tau }(u)/\varphi (u)$.
In virtue of Eq.(\ref{corr2}),
\begin{equation}
\mathrm{Cov}[\hat{\chi}_{\tau }(t),\xi (s)]=0.  \label{corr3}
\end{equation}

In terms of the stochastic process $\zeta (t)$, the BLF function looks like
\begin{equation}
\hat{\chi}_{\tau }(t)=\mathrm{E}[\chi ] +  \sum_{s=\tau}^{+\infty }C(s)\zeta (t-s).
\label{zeta}
\end{equation}

At $s=\tau,$ we obtain 
\begin{equation}
\frac{\Xi _{\tau }(0)^{(\tau)}}{\tau !}=\frac{\sum_{i=1}^{n}c_{i}e^{-%
\alpha _{i}\tau }}{\sum_{i=1}^{n}c_{i}},  \label{l=0}
\end{equation}
and at $l = s - \tau > 0$
\begin{equation}
\frac{\Xi _{\tau }(0)^{(s)}}{s!}=D_{n}^{-1/2}%
\sum_{j=1}^{n-1}e{}^{-\nu _{j}(l-1)}\left( \prod_{i\neq j}^{n-1}(e^{-\nu
_{j}}-e^{-\nu _{i}})\right) ^{-1}\sum_{i=1}^{n}c_{i}e^{-\alpha _{i}\tau
}\prod_{k\neq i}^{n}(e^{-}{}^{\nu _{j}}-e^{-\alpha _{k}}).  \label{lgt0}
\end{equation}

The last two equations complete solution of the BLF problem for the case when the
correlation function is a superposition of the exponent functions (\ref
{corr}). The $n(n-1)$ terms in the right side of Eq.(\ref{lgt0}) are not all 
positive definite.

\begin{table}[tbp]
\caption{ Parameters $d_{i}$ and $\alpha _{i}$ entering the fit of the
autocorrelation function (\ref{corr}) of the reduced volatility, parameters $%
\nu _{i}$ which determine the roots of equation $P_{n-1}(\phi _{0})=0,$ and
parameters $c_{i}$ which determine the additive representation (\ref{phi
additive})\ of the function $\varphi (u)$. The value of $D_{4}$ represents
the normalization constant according to Eq.(\ref{parameters}). }
\label{lab1}
\begin{center}
\begin{tabular}{llllllll}
\hline\hline
$d_{1}$ & $0.40$ & $\alpha _{1}$ & $+\infty$ & $\nu _{1}$ & $.002257$ & $%
c_{1}$ & $.606241$ \\ 
$d_{2}$ & $0.05$ & $\alpha _{2}$ & $1/20$ & $\nu _{2}$ & $.012107$ & $c_{2}$
& $.038233 $ \\ 
$d_{3}$ & $0.03$ & $\alpha _{3}$ & $1/250$ & $\nu _{3}$ & $.125302$ & $c_{3}$
& $.007857$ \\ 
$d_{4}$ & $0.05$ & $\alpha _{4}$ & $1/1000$ & $D_{4}$ & $.435341$ & $c_{4}$
& $.007473$ \\ \hline\hline
\end{tabular}
\end{center}
\par
\vspace*{13pt}
\end{table}

\section{Parameters of BLF volatility function for Dow Jones 30
Industrial Average}

Let us apply the BLF method to forecasting the volatility for the DJIA. The
daily returns are defined by Eq.(\ref{defreturn}) where $S(t)$ are the DJIA
index close values, the volatility equals $\eta (t)=|\xi (t)|,$ and $\chi (t)
$ is defined by Eq.(\ref{mod}). In Fig.1, we show the empirical values of
the correlation coefficients \textrm{Corr}$[\eta (t),\eta (0)],$ \textrm{Corr%
}$[\eta (t),\xi (0)]$ and \textrm{Corr}$[\chi (t),\chi (0)],$ \textrm{Corr}$%
[\chi (t),\xi (0)]$ and the exponential fit of the correlation coefficient 
\textrm{Corr}$[\chi (t),\chi (0)]$ versus the time lag $t.$ Let us remind
that \textrm{Corr}$[A,B]=$\textrm{Var}$[A,B]/\sqrt{\mathrm{E}[A^{2}]\mathrm{E%
}[B^{2}]}$ and $-1\leq \mathrm{Corr}[A,B]\leq 1.$ 
To calculate $\chi (t),$ we run the summation
over $s$ in Eq.(\ref{mod}) from $0$ to $250$ and use the empirical
correlation function of $\eta (t)$ and $\xi (s)$ without additional
smearing. Up to $t=250$,  the correlation coefficient \textrm{Corr}$[\chi (t),\xi (0)]\ $
is less noisy as compared with other correlators.
The parameters $d_{i}$ and 
$\alpha _{i\text{ }}$are listed in Table 1. The equation $P_{3}(\phi )=0$
determines the parameters $\nu _{i}$ ($i=1,2,3$) according to Eq.(\ref{poly}%
). Using Eqs.(\ref{parameters}), we find the values of $D_{4}$ and $c_{i}$,
which we also show in Table 1. One can check that Eq.(\ref{di}) is
satisfied. The correlation coefficient \textrm{Corr}$[\chi (t),\chi (0)]$
drops from $0.53$ ($=\sum_{i=1}^{4}d_{i}$) to $0.13$ ($=\sum_{i=2}^{4}d_{i}$%
) when $t$ changes by one unit from $t=0$ to $t=1$ . The value of $\alpha
_{1}$ is therefore large and can be fixed by considering high-frequency data
only. The results shown on Fig. 1 and in Table 1 and Table 2 are obtained
for $\alpha _{1}=+\infty .$ The values of $1/\alpha _{i}$ for $i=2,3,4$
equal to about one month, one year, and four calendar years, respectively.

The weight coefficients $\Xi _{\tau }(0)^{(s)}$ can be found with the
use of Eq.(\ref{l=0}) at $s = \tau$ and Eq.(\ref{lgt0}) at $s > \tau$. We show values
of the weight coefficients divided by $s!$ in Table 2 for $%
l = s - \tau = 0,1,2,$ and $10$ and $\tau =1,2,10,$ and $100.$

\begin{table}[tbp]
\caption{ The weight coefficients (\ref{l=0}) and (\ref{lgt0}) of the BLF
function for some values of the parameters $l\ $and $\tau .$
Here, $\tau $ is the forecast horizon, $l = s - \tau$ is the number of trading day entering
the predictor function starting from the most recent day. }
\label{lab2}
\begin{center}
\begin{tabular}{lllll}
\hline\hline
$l$ & $\tau =1$ & $\tau =2$ & $\tau =10$ & $\tau =100$ \\ \hline
$0 $ & $.07830$ & $.07555$ & $.05780$ & $.01862$ \\ 
$1$ & $.06942$ & $.06702$ & $.05150$ & $.01710$ \\ 
$2$ & $.06158$ & $.05949$ & $.04594$ & $.01575$ \\ 
$10$ & $.02432$ & $.02366$ & $.01941$ & $.00916$ \\ 
\hline\hline
\end{tabular}
\end{center}
\par
\vspace*{13pt}
\end{table}

The BLF volatility function looks like
\begin{equation}
\hat{\eta}_{\tau }(t)=\mathrm{E}[\chi ]+\sum_{s=\tau}^{+ \infty }\frac{\Xi _{\tau
}(0)^{(s)}}{s!}(\chi (t - s)-\mathrm{E}[\chi
])+\sum_{s=\tau}^{+ \infty }\mathrm{Cov}[\eta (0),\xi (-s)]\mathrm{Var}%
^{-1}[\xi ]\xi (t-s)  \label{final}
\end{equation}
where the unknown future returns set qual to zero: $\xi (t-s)\rightarrow 
\mathrm{E}[\xi (t-s)]=0$ for $0\leq s<\tau $.

Using Eqs.(\ref{corr1}), (\ref{corr2}), and (\ref{corr3}), one gets
\begin{equation}
\mathrm{E}[\left( \hat{\eta}_{\tau }(t)-\eta (t)\right) ^{2}]=\mathrm{E}%
[\left( \hat{\chi}_{\tau }(t)-\chi (t)\right) ^{2}]+\sum_{s=0}^{\tau -1}%
\mathrm{Cov}[\eta (0),\xi (-s)]^{2}\mathrm{Var}^{-1}[\xi ],  \label{err}
\end{equation}
so the minimization of the $\hat{\chi}_{\tau }(t)$ error according to
Eq.(\ref{xihat}) is equivalent to the minimization of the $\hat{%
\eta}_{\tau }(t)$ error. Using decompositions (\ref{firepr}) and (\ref{zeta}), the 
$\hat{\chi}_{\tau }(t)$ error can be evaluated as
\begin{equation}
\mathrm{E}[\left( \hat{\chi}_{\tau }(t)-\chi (t)\right) ^{2}]=\mathrm{E}%
[\chi ^{2}]\sum_{s=0}^{\tau -1}C^{2}(s)=\mathrm{E}[\chi
^{2}]\sum_{i=1}^{n}\sum_{k=1}^{n}c_{i}c_{k}\frac{1-e^{-(\alpha _{i}+\alpha
_{k})\tau }}{1-e^{-(\alpha _{i}+\alpha _{k})}}.  \label{theorerror}
\end{equation}
(There is a misprint in Eq.(10.2) of Ref. \cite{portenko}) At $\tau \rightarrow
+ \infty $, $\mathrm{E}[\left( \hat{\chi}_{\tau }(t)-\chi (t)\right) ^{2}]
\rightarrow \mathrm{E}%
[\chi ^{2}]\sum_{i=1}^{n}d_{i}=\mathrm{Var}[\chi ]$, in agreement with the
fact that $\hat{\chi}_{\tau }(t)\rightarrow \mathrm{E}[\chi ].$ The arguments
of such a kind do not apply to the variance estimator  $v(t)=|\xi (t)|^{2}$, since  
$\mathrm{Var}[v] = \infty$ due to the power-law tails of the return distributions \cite{para}.

Nonlinear models for volatility forecasting \cite{mik,mikw}, which take into
account besides the volatility clustering and leverage effect also heavy
tails of the returns distributions and the approximate scaling, represent an
alternative class of the stochastic volatility models. The efficiency of
such models can be tested in general using Monte Carlo simulations and/or
backtests over historical data. The approach of Refs. \cite{mik,mikw} is
more general, since it allows a calculation of the probability density
function of the volatility. The BLF method predicts the average volatility
only. It can, however, be extended to forecasting $|\xi |^{a}$ for
arbitrary $0<a$ such that \textrm{E}$[|\xi |^{2a}]<\infty .$ If all moments 
\textrm{E}$[|\xi |^{a}]$ of the future distribution are known, the
reconstruction of the probability density function of the volatility must be
possible within the BLF method also. 

\section{Conclusion}

The BLF problem for a stationary stochastic process was formulated in 1941 year 
by Kolmogorov \cite{kolm} and later by Wiener \cite{wien}. A modern review of
the BLF methods can be found in Ref. \cite{portenko}. In this paper, we reported
an explicit analytical solution of the BLF problem for practically important
case when the autocorrelation function represents a superposition of
exponential functions. The autocorrelation function of the volatility in a
financial time series is known to be fitted well by such a superposition. We applied
the obtained results to construct the BLF volatility function for the DJIA.

The popular autoregressive conditional heteroskedasticity (ARCH) models of
time dependent volatility, proposed by Engle \cite{engle} (for a review see 
\cite{arch1,arch2,arch3}), describe the variance $\sigma (t)^{2}$ as a linear
function of the past observables. The ARCH models are conceptually very
close to the BLF approach. Eq.(\ref{final}) expresses the forecasting
volatility also as a linear function of the past volatility and past returns. Eq.(%
\ref{final}) gives, however, the best linear forecast with the proved
smallest forecasting error (\ref{theorerror}). The weight coefficients $\Xi
_{\tau }(0)^{(s)}$ allow to evaluate the magnitude and number of terms
needed for the ARCH models to quantify future variance with sufficiently
good precision. The ARCH models receive an additional support and more
general framework through the BLF formula (\ref{final}).

The accurate estimates of the future volatility are important for risk
management and options pricing. The BLF formula (\ref{final}) represents an
interest as the proved most accurate estimate in the class of all linear
functions of the past volatility and past returns.

\begin{acknowledgments}
The author wishes to thank E. Alessio and V. Frappietro for several useful
discussions and the Dow Jones Global Indexes for providing 
the DJIA historical quotes. This work has been supported in part by Federal Program
of the Russian Ministry of Industry, Science and Technology 
No. 40.052.1.1.1112. 
\end{acknowledgments}

\end{document}